\documentclass[12pt,preprint]{aastex}

\usepackage{emulateapj5}

\newcommand{\msol}{$M_{\odot}$}

\newcommand{\e}[1]{$10^{#1}$}
\newcommand{\cm}[1]{\,cm$^{#1}$}
\newcommand{\ee}[1]{$\times 10^{#1}$}

\newcommand{\erg}{~erg\,cm$^{-2}$\,s$^{-1}$}

\newcommand{\ps}{1WGA~J1713.4--3949}

\begin{document}

\title{X-ray observations of the compact central object in supernova
remnant G347.3--0.5}

\author{J. S. Lazendic\altaffilmark{1}, P. O. Slane\altaffilmark{1}, 
B. M. Gaensler\altaffilmark{1}, P. P. Plucinsky\altaffilmark{1},\\ 
J. P. Hughes\altaffilmark{2}, D. K. Galloway\altaffilmark{3} and
 F. Crawford\altaffilmark{4}}

\altaffiltext{1}{Harvard-Smithsonian Center for Astrophysics, 
60 Garden Street, Cambridge, MA 02138, jlazendic@cfa.harvard.edu}
\altaffiltext{2}{Department of Physics and Astronomy, Rutgers University, 
136 Frelinghuysen Road, Piscataway, NJ 08854}
\altaffiltext{3}{Center for Space Research, Massachusetts Institute of
Technology, Cambridge, MA 02139}
\altaffiltext{4}{Department of Physics, Haverford College, 
Haverford, PA 19041}

\vfil

\begin{abstract}

We present {\em Chandra, XMM-Newton} and {\em RXTE} observations of
\ps, a compact source at the center of the galactic supernova remnant
(SNR) G347.3--0.5. The X-ray spectrum of the source is well-fitted by
the sum of  a blackbody component with a temperature of $\sim 0.4$~keV
plus a power law component with photon index $\sim 4$.  We found no
pulsations down to 4\% in the 0.01--0.16 Hz range and down to 25\% in
the 0.01-128 Hz range. This source resembles  other compact central
objects (CCOs) in SNRs, and we suggest that \ps\ is the associated
neutron star for G347.3--0.5. We also measured the properties of the
adjacent radio pulsar PSR J1713--3945 with a 392~ms period and show
that it is not associated with \ps\ nor, most probably, with SNR
G347.3--0.5 as well.

\end{abstract}


\section{INTRODUCTION}

For many years it has been well known that pulsars --- compact remains
of supernova explosions --- are rapidly rotating,  highly magnetized
neutron stars (NSs). However, sources  discovered in recent  years
have shown that NSs also appear to come in other varieties: even more
highly magnetized objects with slow spins, such as anomalous X-ray
pulsars (AXPs) and soft gamma-ray repeaters (SGRs)
\citep[e.g.,][]{mereghetti98,thompson00}, and more poorly understood
objects near the centers of SNRs with detectable X-ray flux, but no
optical/radio counterparts or signs of rotation, so-called compact
central objects \citep[CCOs;][]{pavlov02}.  Their spectra are very
soft, with blackbody temperatures $\sim 0.4$~keV,  often requiring
additional power law component with photon indices of $\sim 4$
\citep{pavlov02}. In this article we address the nature of \ps, a
compact source located at the center of the SNR G347.3-0.5 (RX
J1713.7$-$3946). Early {\em ROSAT}  observations identified  two point
sources within the boundaries of the SNR  shell \citep[][; see also
Figure~\ref{fig1}]{pfeffermann96}.  The source further away from the
center of the remnant, 1WGA~J1714.4$-$3945, was determined to be of
stellar origin \citep{pfeffermann96}.  For the other source,
1WGA~J1713.4$-$3949, located at the geometrical center of the SNR, no
optical counterpart has been found within 10\arcsec\ of the {\em
ROSAT} position \citep{slane99}, and the source  has correspondingly
been suggested to be a candidate for an associated  neutron star.
While {\em ASCA} observations were not able to provide useful limits
to the  pulsations in the X-ray band \citep{slane99}, radio pulsations
were  detected with a  392~ms period  within a 7\arcmin\ radius region
towards  this source  \citep[PSR J1713--3945][]{crawford02}.

\ps\ has been observed with the {\em Chandra X-ray Observatory} as a
part of a project to study nonthermal radio and X-ray emission from
the SNR shell \citep{lazendic03}.   In this paper we present a
detailed spectral and timing analysis of \ps\ using our {\em Chandra}
data, as well as archival data from the {\em XMM-Newton {\rm and}
RXTE} satellites.   We also present properties of  the radio pulsar
PSR J1713--3945, which imply it is not related to \ps\ or SNR
G347.3--0.5.

\section{OBSERVATIONS}

Two fields towards G347.3$-$0.5 were observed on 2000 
July 25 with the Advanced CCD
Imaging Spectrometer (ACIS) detector on board the {\em Chandra X-ray
Observatory} (ObsID 736 and 737).  The ACIS-I detector for ObsID 736 
was positioned at the
bright northwestern rim of the SNR and the roll angle was such that
\ps\  fell on the S1 chip (see Figure~\ref{fig1}). Data were taken in
timed exposure mode with an integration time of 3.2~s  per frame.  The
fully processed Level 2 data were reduced using standard threads in
the  {\tt CIAO} software package version 2.2.1.  The effective
exposure time after data processing  was 29.6~ks.
The exposure-corrected {\em Chandra} image of the SNR and the compact
source \ps\ is shown in Figure~\ref{fig1}, with the {\em ROSAT}
contours  overlaid. The image was binned in $4\farcs 6 \times 4\farcs
6$ pixels and smoothed with a Gaussian filter with a FWHM of
2\arcsec. The image of the compact source is elongated because its
location is well off-axis ($\sim$ 23\arcmin) from the aim point. 

We obtained archival {\em XMM} data  of \ps\ carried out   on 2001
March 2  with a 15~ks  exposure  \citep[ObsID 0093670501; see
also][]{cassam03}.  The European Photon Imaging Camera (EPIC)-pn
camera was operated in Extended  Full Frame mode with time resolution
of 199~ms, while the two EPIC-MOS cameras were operated in Full Frame
mode with a time resolution of 2.6~s. The EPIC data reduction was
performed using the {\tt SAS} software package version 5.4.1.  The
event files used for analysis  were created from observational data
files (ODFs)  using the {\tt SAS} tasks {\tt epchain} and {\tt
emchain}.  The net exposure time after filtering event files for good
time intervals was 9.2~ks in the pn camera and 14.1~ks in each MOS
camera.

We used archival data of SNR G347.3--0.5 \citep{pannuti03}   from the
Proportional Counter Array (PCA) aboard the {\em RXTE} satellite,
which is sensitive to X-ray photons in the energy band 2--60 keV and
has a field of view with radius $\approx1$\arcdeg. While the PCA has
no imaging capability, it has  a large effective area (6000~cm$^2$)
which makes it suitable for  searching for faint X-ray pulsations.
Observations of G347.3--0.5 were made in 7 separate intervals between
1999 June 14--24, with a total exposure time of 100~ks.  Analysis of
the {\it RXTE}\/ data was carried out using {\tt LHEASOFT} software
version 5.2. 

A 392-ms radio pulsar, PSR J1713$-$3945, was discovered in a 20 cm
targeted search for pulsations in SNR G347.3--0.5 using the Parkes
telescope and multibeam receiver. The details of the search and
discovery are outlined in \citet{crawford02}. The pulsar was
subsequently timed with the multibeam receiver at Parkes to obtain
astrometric and spin parameters and to determine whether the pulsar
was associated with \ps.


\section{ANALYSIS AND RESULTS}
  
\subsection{Spectral Analysis}

For the spectral analysis of the {\em Chandra} data, we extracted the
source counts from  a $37\arcsec \times 61\arcsec$ elliptical region
around the compact source, and for  the background  region we used a
$2\arcmin \times 3\arcmin$ elliptical region north of  the source. We
obtained $8454\pm103$ background-subtracted counts  for the source and
the spectrum was grouped to contain at  least 25 counts per bin.
Residuals from fitting standard continuum models to the source
spectrum (blackbody, bremsstrahlung and power law) revealed an
apparent change in the gain calibration on chip S1 that results in a
spurious absorption feature  around 2~keV. We extracted data from the
on-board calibration source taken before and after our observation,
and these show that there is indeed a gain shift for this CCD of $\sim
3.5$\%. However,  we found that the fit parameters  did not change
significantly when the gain was frozen at this 3.5\% value or when it
was left as a free parameter. We thus allowed gain to vary in our
models  (see Table~\ref{tab-fit}).

For spectral analysis of the {\em XMM} data, we obtained $9600\pm217$
background-subtracted counts from the pn camera, and 5$451\pm74$
($5274\pm73$) in the MOS1 (MOS2) camera. In the pn camera, the counts
were extracted from a 27\arcsec\ radius circular region around the
source, and for the background extraction we used a $3\farcm 4 \times
8\farcm 3$ box north from the source position inside the same CCD
chip. Similarly, in the MOS1 (MOS2) camera we used 22\arcsec\
(19\arcsec)  radius circular region for the source and $3\farcm 4$
(2\farcm 2)  radius circular region on the same chip for the
background.  
The spectra were also grouped to
contain at least 25 counts per bin. We performed a joint spectral
analysis of the three EPIC data sets, allowing only the relative
normalization to vary. The results obtained were consistent with those
obtained from  the {\em Chandra} spectrum. We therefore proceeded with
a joint spectral analysis of {\em Chandra} and {\em XMM} spectra.

The results from fitting a few typical neutron star  models are
summarized in Table~\ref{tab-fit} and a fit example is given in
Figure~\ref{fig2}. Fits with the bremsstrahlung and blackbody 
 models gave equally acceptable results, while
the power law model gave a somewhat higher reduced $\chi^{2}$.  The
steep photon index of $\Gamma \sim 4$ excludes the possibility that
\ps\ is a background active galactic nucleus (AGN), which typically
have photon indices of 1.2--2.2 \citep{turner86}. The best-fit column
density ($N_H$) in the power law model of $\sim 11$\ee{21}\cm{-2}  is
also somewhat higher than the value of $8 \times
10^{21}$\cm{-2} found towards the SNR \citep{lazendic03}.  On the
other hand, a blackbody model yields $N_H \sim
4$\ee{21}\cm{-2}, which is a lower value than found towards the
SNR. The blackbody fit is improved by adding a power law component,  
 which also results in an $N_H$ value consistent
with that  found towards the SNR. In this composite spectrum the
unabsorbed source luminosity is  $L_{X} (0.5-10.0~{\rm keV}) \sim 6
\times 10^{34}$ erg s$^{-1}$,  adopting a distance of $\sim 6$~kpc to
the SNR \citep{slane99}. The blackbody normalization yields a
relatively small emitting area  with radius $R \sim 2.4~D_{6kpc}$~km,
which is significantly smaller than the canonical value for the NS
radius of 10~km. This could imply that most of the thermal emission
originates from the polar caps on the NS surface, while the thermal
component from the rest of the NS surface is too faint to be detected. 

The small inferred radius may be due, at least partly, to
oversimplification of the thermal emission from the NS by the
blackbody model.  More appropriate models that take into account the
NS atmosphere (NSA) yield  lower NS temperatures, which in turn
correspond to  larger NS radii \citep[e.g.,][]{zavlin96}. To
investigate this, we fit our spectra  with two NSA models. The
 NSA model by \citet{gansicke02} is calculated for weakly 
magnetic NS  ($B \le 10^{10}$~G) and  considers 
hydrogen, solar or iron compositions of the atmosphere.
The NSA model by \citet{pavlov95} assumes the  standard
magnetic field strength ($B= 10^{12}$~G) and hydrogen atmosphere.
 Both models assume a standard NS mass of 1.4\msol.  Both
hydrogen models gave acceptable residuals and yielded emitting region
sizes close to 10~km, with corresponding reductions in the NS
temperature. For either model, however, the derived temperature is
 above the expected surface temperature for a cooling neutron star
more than several hundred years old.  Model atmospheres with solar
abundance or iron composition gave unacceptable fits predicting broad
absorption features not present in the spectra. 


\subsection{Timing Analysis}

We used a Fast-Fourier Transform (FFT) to search the X-ray data  for
pulsations from \ps,  but found no pulsed signal; we list  upper limits
on the pulsed fraction and the search  parameters in
Table~\ref{tab-timing}. The upper limit from {\em RXTE} data is least
constraining because \ps\ only accounts for 0.6\%  of the total flux
in the {\em RXTE}'s large field of view.

To derive the timing properties of the radio pulsar PSR J1713-3945 the
{\tt TEMPO} timing package\footnote{http://pulsar.princeton.edu/tempo}
was used on a total of 27 times-of-arrival (TOAs) with the JPL DE200
planetary ephemeris \citep{standish90}. The {\tt TEMPO} package
converts each TOA to the solar system barycenter then refines the
timing parameters by minimizing residuals between observed and model
TOAs over the observation span. The resulting timing parameters are
presented in Table~\ref{tab-pulsar}. The pulsar's dispersion measure
(DM) of 337~pc~cm$^{-3}$ yields an estimated distance of 4.3~kpc using
the revised DM-distance model of \citet{cordes02}.

\section{DISCUSSION}

The position of the \ps\ determined from the {\em XMM} data is
R.A~(J2000) $17^{\rm h}~13^{\rm m}~28^{\rm s}.4$,  Decl. (J2000)
$-39\degr 49\arcmin 54\farcs 5$ (with an uncertainty  of $\sim
6\arcsec$).  Using our 20 cm radio data of SNR G347.3--0.5 obtained
with Australia Telescope Compact Array \citep{lazendic03}, we found no
radio counterpart for \ps, and derive a 5$\sigma$ upper limit for a
point source flux of 3~mJy. This is similar to the limit derived
towards the CCO in Puppis A \citep{gaensler00}.  There is no apparent
optical emission at the location of \ps\ in the  Digitized Sky
Survey\footnote{DSS was produced at the Space Telescope Science
Institute under U.S. Government grant NAG W-2166.} image, which has an upper
limit of $V> 17$~mag.

\begin{figure}
\plotone{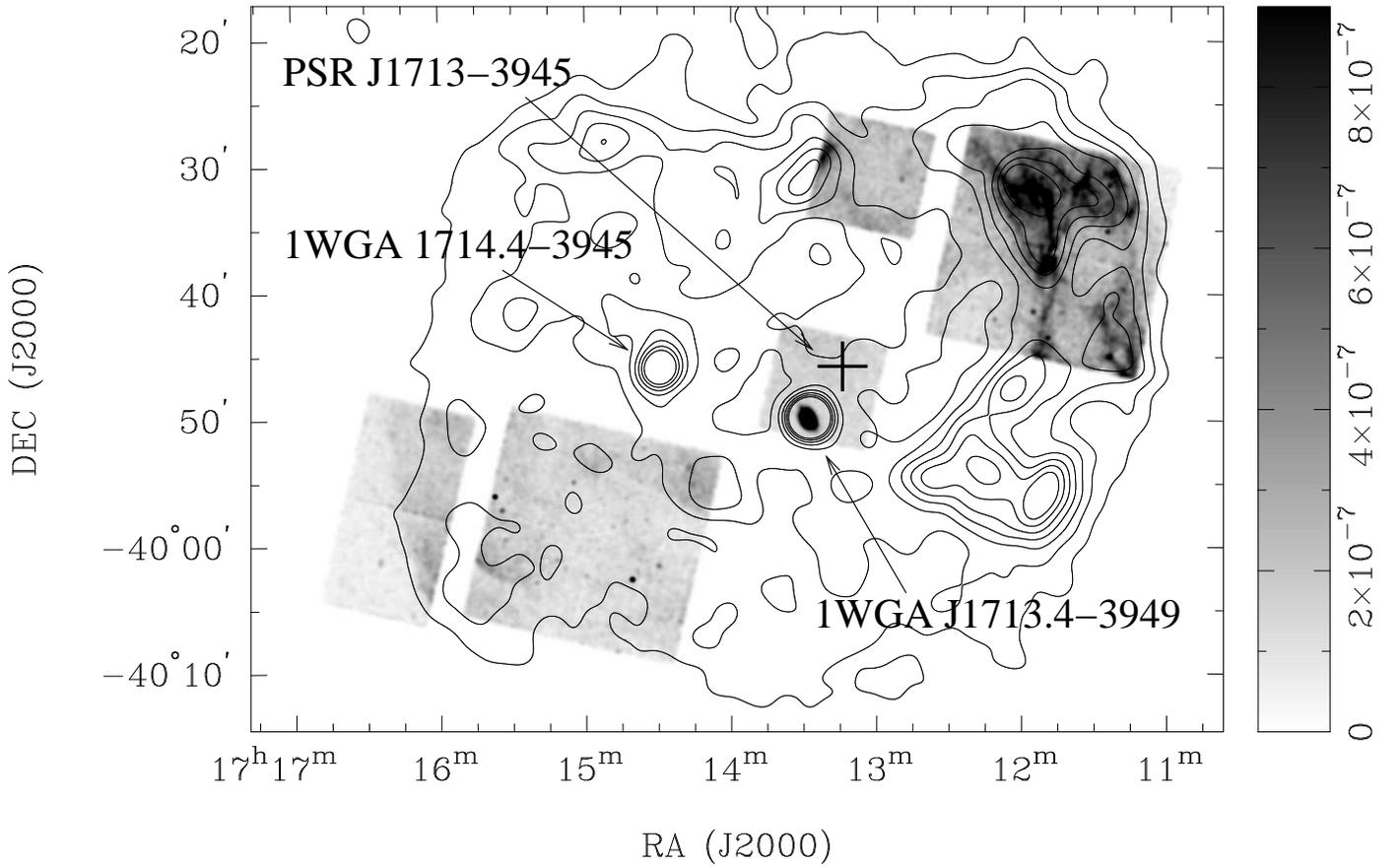}
\caption{The 0.5-10.0 keV band {\em Chandra} greyscale image of
 the two fields observed towards the SNR G347.3--0.5 
\citep[for more details see][]{lazendic03}. 
To show the whole extent of
the remnant, the {\em Chandra} image is overlaid with the 
{\em ROSAT} PSPC contours \citep{slane99}. Two point sources
 detected with {\em ROSAT} observations are labeled, as well as the
location of the radio pulsar PSR J1713$-$3945.}
\label{fig1}
\end{figure}

\begin{figure}
\plotone{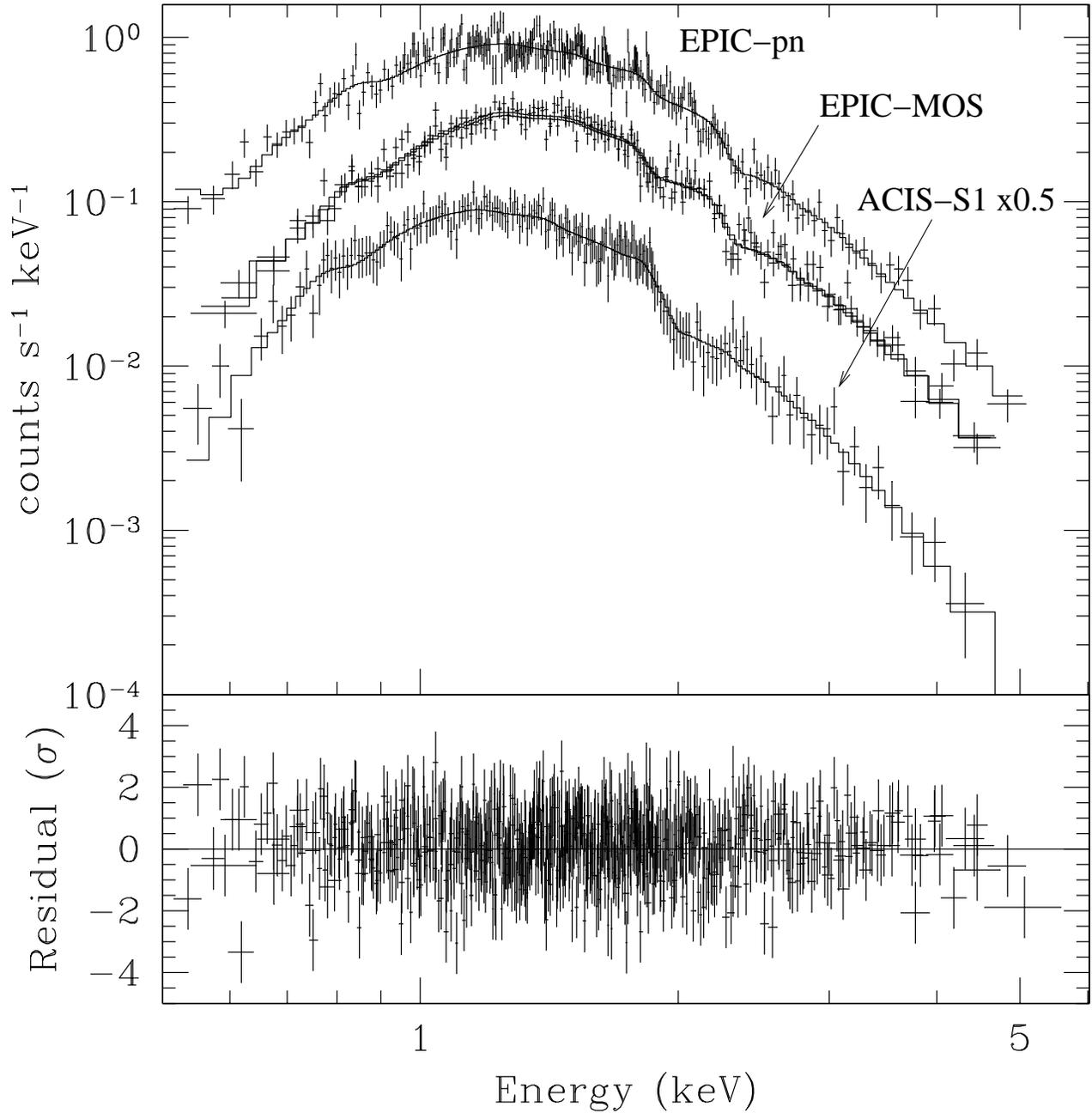}
\caption{ACIS and EPIC spectra from the compact source \ps\ 
and residuals for blackbody+power law model. The fit parameters are
listed in  Table~\ref{tab-fit}.} 
\label{fig2}
\end{figure}

While the distance to the radio pulsar PSR J1713--3945 is  broadly
consistent with the SNR distance of $6 \pm 1$ kpc  estimated by
\citet{slane99}, the pulsar is likely to be quite old ($\tau_{c} \sim
1.1$~Myr, compared to $\tau < 40$~kyr for SNR G347.3--0.5) and does
not have sufficient spin-down luminosity ($\dot{E} \sim 3.7 \times
10^{33}$~erg~s$^{-1}$) to power the observed X-ray flux from \ps\
($L_X \sim 6 \times 10^{34}$~erg~s$^{-1}$).  The newly determined
radio timing position of the pulsar is $\sim 4\arcmin$ in declination
away from the position of \ps\ (see Figure~\ref{fig1}).  PSR
J1713$-$3945 is therefore spatially coincident with  SNR G347.3--0.5
by chance, and there is likely no  physical association between the
two systems.  We detected no X-ray source at the position of PSR
J1713--3945 in {\em Chandra} or {\em XMM} data. Adopting 
$N_H \sim 5\times 10^{21}$\cm{-3} (by scaling the value for the SNR 
to the 4.3~kpc distance of this pulsar), we derive an upper limit
 on the luminosity in the 0.5--10.0~keV band of 
$2 \times 10^{31}$~erg~s$^{-1}$ for 
nonthermal emission with a photon index $\sim 1.5$, and an 
upper limit on the temperature of 82~eV for blackbody emission 
from a 10-km radius NS. 

The lack of radio and optical counterparts for \ps, the absence of
X-ray  pulsations with the current sensitivity and time resolution,
the two-component spectrum and its associated luminosity  are
properties consistent with CCOs \citep{pavlov02}.  

If this CCO is an accretion powered NS, we can consider two scenarios: 
 accretion from a fallback
disk or a low-mass companion.  The fallback disk has been suggested to
form from residual material after the SNR explosion
\citep[e.g.,][]{chevalier89} and  has been used to explain and unify
the properties of various types of NSs \citep{alpar01}.  However,
limits derived from optical and infrared  observations seem to
disfavor this model, at least for some types of NSs
\citep[e.g.,][]{kaplan02}.  X-ray monitoring, in conjunction
with optical and infrared limits, can be used to establish if the NS
is variable and thus likely to be accreting from a   low-mass
companion. Indeed,  variability and a period of 6.4~h 
has been reported for one of the CCOs,
J1617--5102 in SNR RCW 103, on which basis this CCO is claimed to be a
binary  \citep{sanwal02}. Comparing the unabsorbed X-ray flux
of \ps\ obtained from {\em Chandra} and {\em XMM} data  for a
blackbody fit with that of {\em ASCA} data
\citep[$F_{X}^{0}(0.5-10.0)=5.3$\ee{-12}\erg;][]{slane99}, we find no
long-term ($\sim 3$~yr) variability in \ps.

Interpretations of the emission from CCOs as thermal radiation from
the surface of a NS run into problems  due
to high surface temperatures and small emitting areas  derived from
the blackbody model fits. The temperature we derived   from
the blackbody model of 0.4~keV is much higher that the temperature
predicted by the standard cooling model \citep{page96}, while the
temperature derived from the NSA models (0.2~keV) would imply an NS
age of 100--1000 yr, which is  less than the estimated 20--40 kyr age
of the SNR \citep{slane99}. 

In the case that CCOs are rotation powered NSs, the non-detection of
radio pulsations can be explained by unfavorable beaming orientation.
One of the objects,  1E~1207.4--5209 in SNR G296.5+10.0,   that was
originally identified as a CCO, has been found to  have a period of
424~ms with faint pulsed fraction  \citep[7.6\%;][]{zavlin00}. Our
limits on pulsed fraction from \ps\  are consistent with those derived
towards other  CCOs \citep[7--15\%; e.g.,][]{pavlov02}. 
Our limits indicate  that \ps\ is probably not an AXP, which  
are generally found to have spin periods between 6--12~s  and 
pulse fractions of 10--70\% \citep[e.g.,][]{mereghetti98}. 

In summary, the nature of the compact central object in SNR
G347.3--0.5 remains uncertain.  Future observations are planned to
help determine the exact nature of this source.   Pointed {\em
Chandra} observations are needed  to provide  a more accurate position
and to enable a search for an optical counterpart, while deeper  timing
observations can search for periods substantially
shorter than available from the current data.

\acknowledgements

We thank Peter Woods for useful discussions and Slava Zavlin for
providing the files for the magnetic neutron star atmosphere.
This work was supported in part by NASA contract NAS8--39073 (POS), 
grant GO0--1123X (JSL) and grant GO2--3080B (JPH).



\clearpage

\begin{deluxetable}{lccccccc}
\scriptsize
\tablecaption{Parameters of the joint spectral fits to 
the {\em Chandra} and {\em XMM} spectra of the compact 
source \ps\ with different models. \label{tab-fit}}
\tablewidth{0pt}
\tablehead{
\colhead{Model} & \colhead{$N_H$} & \colhead{$kT$} & \colhead{$\Gamma$} 
 & \colhead{$R_{\infty}\tablenotemark{a}$} & \colhead{$\chi^{2}_{red}$/dof} & \colhead{Gain\tablenotemark{b}} & \colhead{$F_{X(0.5-10.0)}^{0\tablenotemark{c}}$} \\
\colhead{ } & \colhead{(\e{21} \cm{-2})} & \colhead{(keV)} 
& \colhead{} & \colhead{(km) } & \colhead{ } & \colhead{(\%)} & \colhead{(\erg)}
}

\startdata

BREMSS       & 7.0$^{+0.1}_{-0.2}$ & 0.92$_{-0.02}^{+0.02}$ & --
             & -- & 1.07/675 & 5.1$_{-0.2}^{+0.5}$ & 9.4\ee{-12} \\

POW           & 10.7$_{-0.2}^{+0.4}$ & -- & 4.2$_{-0.1}^{+0.1}$ 
              & -- & 1.44/675 & 4.6$_{-0.5}^{+0.8}$ 
              & 3.2\ee{-11}  \\

BB           & 3.6$_{-0.1}^{+0.1}$ & 0.40$_{-0.01}^{+0.01}$ & -- 
             & 2.5$_{-0.1}^{+0.1}$
             & 1.25/675 & 6.0$_{-0.6}^{+0.5}$ & 4.5\ee{-12} \\

POW+BB       & 8.0$_{-0.6}^{+0.9}$ & 0.38$_{-0.02}^{+0.01}$ 
             & 3.9$_{-0.2}^{+0.3}$ &  2.4$_{-0.4}^{+0.6}$
             & 1.05/670 & 7.1$_{-0.8}^{+0.9}$ & 1.4(0.3)\ee{-11} \\

POW+NSA(G)\tablenotemark{d}  & 7.3$_{-0.4}^{+0.6}$ & 0.18$_{-0.01}^{+0.02}$ 
             & 3.7$_{-0.3}^{+0.8}$ & 13.0$_{-3.5}^{+1.0}$
             & 1.07/670 & 5.5$_{-0.5}^{+0.6}$ & 1.1(0.4)\ee{-11} \\

POW+NSA(Z)\tablenotemark{e}  & 8.1$_{-0.9}^{+1.1}$ & 0.26$_{-0.06}^{+0.03}$ 
             & 3.8$_{-0.5}^{+0.2}$ & 15.8$_{-8.4}^{+4.2}$ 
             & 1.22/667 & 5.1$_{-0.5}^{+0.5}$ & 1.5(0.3)\ee{-11} \\

\enddata

\noindent
\tablenotetext{a}{radius of the emitting region at infinity assuming
the source distance of 6~kpc}
\tablenotetext{b}{a linear gain shift included for fitting  the {\em
Chandra} data}
\tablenotetext{c}{unabsorbed flux --- for the two component models 
the flux of the second component alone is given in parentheses}
\tablenotetext{d}{NSA model by \citet{gansicke02} for weak magnetic
field of $\le 10^{10}$~G}
\tablenotetext{e}{NSA model by \citet{pavlov95} for standard magnetic 
field of $10^{12}$~G} 
\end{deluxetable}


\begin{deluxetable}{lclr}
\scriptsize
\tablecaption{Parameters and results of pulsation search from \ps\
using FFT. \label{tab-timing}}
\tablewidth{0pt}
\tablehead{
\colhead{Detector} & \colhead{Time} & \colhead{Frequency range}  
& \colhead{$f_p$ limit\tablenotemark{a}} \\
\colhead{ } & \colhead{resolution} & \colhead{searched} &  
\colhead{} }
\startdata
ACIS-S1   & 3.2~s    & 0.01--0.16 Hz & 12\% \\
EPIC-MOS\tablenotemark{b}  & 2.6~s    & 0.01--0.19 Hz & 12\% \\
EPIC-pn   & 199~ms   & 0.01--2.5 Hz & 4\% \\
PCA\tablenotemark{c} & 4~ms & 0.01--128 Hz & 25\% \\ 
\enddata
\tablenotetext{a}{fractional pulsation upper limit to 99\% confidence
level calculated according to \citet{vaughan94}}
\tablenotetext{b}{from combined MOS1 and MOS2 data}
\tablenotetext{c}{calculated from FFT of two separate
intervals of $\sim 3$ days of data with gaps} 
\end{deluxetable}



\begin{deluxetable}{ll}
\scriptsize
\tablecaption{Properties of the radio pulsar PSR J1713$-$3945 detected
towards G347.3--0.5. \label{tab-pulsar}}
\tablewidth{0pt}
\tablehead{
\colhead{Parameter} & \colhead{Value\tablenotemark{a}}
}
\startdata
R.A. (J2000)       & $17^{\rm h}~13^{\rm m}~14^{\rm s}_{.}23(3)$ \\
Decl. (J2000)      & $-39\degr 45\arcmin 47\farcs 5(14)$ \\
Period, $P$ (ms)                   &    392.451769234(8) \\
Period derivative, $\dot{P}$ ($\times 10^{-15}$) & 5.7380(5) \\
Dispersion Measure, DM (pc cm$^{-3}$)  & 337(3) \\
Epoch of period (MJD)                  & 52165.0 \\
rms residual (ms)                      & 1.1 \\
Number of TOAs                         & 27 \\
Timing span (Days)                     & 1200 \\
Characteristic age\tablenotemark{b}, $\tau_{c}$ (Myr)   & 1.08 \\
Surface magnetic field\tablenotemark{c}, $B$ ($\times 10^{12}$ G) & 1.52 \\
Spin-down luminosity\tablenotemark{d}, $\dot{E}$ (erg s$^{-1}$) & 3.74\ee{33} \\
\enddata

\tablenotetext{a}{Values in parentheses represent the 1$\sigma$ uncertainty
in the least significant digit quoted.}
\tablenotetext{b}{$\tau_{c} \equiv P / 2 \dot{P}$}
\tablenotetext{c}{$B \equiv 3.2 \times 10^{19} (P \dot{P})^{1/2}$~G,
where $P$ is in seconds}
\tablenotetext{d}{$\dot{E} \equiv 4 \pi^{2} I \dot{P} / P^{3}$, with
$I = 10^{45}$~g~cm$^{2}$ assumed} 

\end{deluxetable}


\clearpage

\clearpage

\end{document}